\documentclass[twocolumn]{aastex63}
\accepted{in ApJ 27.04.2020}
\shorttitle{IMF stochasticity and exoplanets}
\shortauthors{Bottrill, Haigh, Hole, Theakston, Allen, Grimmett \& Parker}

\begin{document}

\title{Exoplanet detection and its dependence on stochastic sampling of the stellar Initial Mass Function}

\correspondingauthor{Richard Parker}
\email{R.Parker@sheffield.ac.uk}

\author{Amy L. Bottrill}
\affiliation{Department of Physics and Astronomy, The University of Sheffield, Hicks Building, Hounsfield Road, Sheffield, S3 7RH, UK}
\nocollaboration{1}
\author{Molly E. Haigh}
\affiliation{Department of Physics and Astronomy, The University of Sheffield, Hicks Building, Hounsfield Road, Sheffield, S3 7RH, UK}
\nocollaboration{1}
\author{Madeleine R. A. Hole}
\affiliation{Department of Physics and Astronomy, The University of Sheffield, Hicks Building, Hounsfield Road, Sheffield, S3 7RH, UK}
\nocollaboration{1}
\author{Sarah C. M. Theakston}
\affiliation{Department of Physics and Astronomy, The University of Sheffield, Hicks Building, Hounsfield Road, Sheffield, S3 7RH, UK}
\nocollaboration{1}
\author{Rosa B. Allen}
\affiliation{King Edward VII High School, Glossop Road, Sheffield, S10 2PW, UK}
\affiliation{Department of Physics and Astronomy, The University of Sheffield, Hicks Building, Hounsfield Road, Sheffield, S3 7RH, UK}
\nocollaboration{1}
\author{Liam P. Grimmett}
\affiliation{Department of Physics and Astronomy, The University of Sheffield, Hicks Building, Hounsfield Road, Sheffield, S3 7RH, UK}
\nocollaboration{1}
\author[0000-0002-1474-7848]{Richard J. Parker}
\altaffiliation{Royal Society Dorothy Hodgkin Fellow}
\affiliation{Department of Physics and Astronomy, The University of Sheffield, Hicks Building, Hounsfield Road, Sheffield, S3 7RH, UK}
\nocollaboration{1}




\begin{abstract}
Young Moving Groups (YMGs) are close ($<$100\,pc), coherent collections of young ($<$100\,Myr) stars that appear to have formed in the same star-forming molecular cloud. As such we would expect their individual initial mass functions (IMFs) to be similar to other star-forming regions, and by extension the Galactic field. Their close proximity to the Sun and their young ages means that YMGs are promising locations to search for young forming exoplanets. However, due to their low numbers of stars, stochastic sampling of the IMF means their stellar populations could vary significantly. We determine the range of planet-hosting stars (spectral types A, G and M) possible from sampling the IMF multiple times, and find that some YMGs appear deficient in M-dwarfs. We then use these data to show that the expected probability of detecting terrestrial magma ocean planets is highly dependent on the exact numbers of stars produced through stochastic sampling of the IMF.

\end{abstract}


\section{Introduction} \label{sec:intro}

Most stars form in groups (often loosely referred to as clusters) with membership ranging between 10s to $10^4$ stars \citep{Lada03}. It is thought that planets form from circumstellar discs of dust and gas almost immediately after star formation, and certainly during the earliest stages of a pre-main sequence star's life \citep{Haisch01,Brogan15,Richert18}. 

Planets can be directly detected around young ($<$50\,Myr) stars, in part due to the strong internal heat source in forming planets \citep{Chabrier14}. The nearest young stars to the Sun are in so-called `Young Moving Groups'; collections of tens to hundreds of stars of similar ages with coherent proper motion velocities \citep{Zuckerman04,Torres08,Mamajek05,Riedel17,Gagne18a,Gagne18b,Gagne18c,Lee19b}. It is unclear whether these YMGs are the outcome of diffuse (low density), low-mass star formation, or whether they are the remnants of more populous star clusters that are in the process of dissolving into the Galactic field.

Due to their young ages and close proximity ($<$100\,pc), YMGs are ideal locations for detecting young exoplanets. Future telescopes and instrumentation may even be able to detect the formation signatures of terrestrial planets. For example, \citet{Bonati19} recently calculated the probability of detecting magma ocean planets in nearby YMGs with the Extremely Large Telescope (ELT).  Magma ocean planets are forming protoplanets with a molten surface, caused by collisions with planetessimals in the protoplanetary disc \citep[e.g.][]{Benz90,Tonks93,Canup01,Nakajima15,Nakajima20}, and hence directly trace terrestrial planet formation.

For a star-forming region with a given number of stars, the initial mass function predicts the numbers of low-mass ($<$3\,M$_\odot$) stars of different spectral types. However, if star formation creates stars by randomly sampling this IMF, for small total numbers of stars this can translate into very different numbers of stars of a given spectral type \citep[e.g.][]{Parker07}.

This becomes problematic when considering membership probabilities for YMGs. Several of the observed YMGs appear to show a deficit in the number of M-stars, but it is unclear if this is due to stochastic sampling (i.e.\,\,low-number statistics), or incompleteness in observations. Given the constantly improving membership censuses of star-forming regions from $\emph{Gaia}$ and ground-based surveys (e.g.\,\,Gaia-ESO), it is possible that YMGs may currently be `underrepresented', with hitherto undiscovered members. In such a scenario, we would expect membership lists of YMGs to be added to in future, which would also improve the chances of detecting planets. Alternatively, if the IMF is stochastically sampled, this may result in an over- or underproduction of stars of a particular spectral type, even if membership is complete.

In this paper, we determine whether stochastic sampling of the IMF can explain the observed deficit of low-mass M-stars in some nearby YMGs, and then use the calculations in \citet{Bonati19} to determine the probability  of detecting magma ocean planets in these YMGs based on updated observational membership data. We then calculate the range of possible magma ocean detection probabilities assuming a stochastically sampled IMF, for hypothetical YMGs at similar distances to those observed.

The paper is organised as follows. In Section~2 we outline our method, in Section~3 we present our results, and we conclude in Section~4.    

\section{Method}

In this section we describe our Monte Carlo simulations to sample from the stellar initial mass function (IMF) and the procedure to calculate the probability of detecting magma ocean planets in nearby Young Moving Groups (YMGs) following \citet{Bonati19}.


\begin{table*}
  \caption{Observed total numbers of stars in ten nearby YMGs, as well as the number of A-, G- and M-stars within each sample in the dataset compiled by \citet{Gagne18a} and augmented by \citet{Gagne18c}. We also list the distance ($d$), age ($\tau_\star$) and latest observed spectral type in each YMG. The three YMGs for which we will calculate the probabilities for detecting magma ocean planets are listed first.}
  \begin{center}
    \begin{tabular}{cccccccc}
      \hline
YMG & $\tau_\star$ & $d$ & $N_{\rm stars}$ & $N_{\rm A-stars}$ & $N_{\rm G-stars}$ & $N_{\rm M-stars}$ & Latest Sp.~Type\\
& (Myr) & (pc) & & & & &\\
     \hline	
$\beta$~Pic & 23 & 37 & 44 & 3 & 4 & 20 & L7\\
TW~Hyd & 10 & 53 & 22 & 2 & 0 & 16 & M9.5 \\
$\eta$~Cha & 11 & 94 & 16 & 2 & 0 & 12 & M6 \\
\hline
AB~Dor & 150 & 20 & 52 & 5 & 13 & 13 & L8\\
Carina & 45 & 65 &  6 & 0 & 1 & 0 & K3 \\
Tuc Hor & 45 & 48 & 43 & 2 & 11 & 5 & L0 \\
Columba & 42 & 50 & 26 & 5 & 7 & 2 & L1 \\
Coma~Ber & 560 & 85 & 38 & 10 & 8 & 0 & K4.9 \\
32 Ori &  22 & 92 & 35 & 0 & 1 & 28 & M5 \\
$\chi^1$~For & 50 & 99 & 11 & 6 & 2 & 0 & G8\\
       \hline
    \end{tabular}
  \end{center}
  \label{obs_data}
\end{table*}

\subsection{Observational data}

Before conducting our Monte Carlo simulations, we first need to define an observed sample with which to compare the results of our simulations. \citet{Bonati19} compiled a list of members of YMGs from the literature, but in this work we will use a more recent compilation from \citet{Gagne18a,Gagne18c}.

\citet{Gagne18a} produced a comprehensive list of known members of nearby YMGs, and augmented this list with newly discovered candidates \citep{Gagne18c}. We include the new members classified by \citet{Gagne18c} in our analysis. We exclude objects labelled as companion stars (i.e.\,\,the secondary or tertiary member of a multiple system) as we are comparing the observational data to the number of stars produced by sampling the \emph{system} IMF.

In our analysis of the IMFs of YMGs, we include all of the young ($\leq$50\,Myr), nearby ($\leq$100\,pc) YMGs in the literature, and we also include two older groups -- AB~Dor and Coma Bernices. However, their advanced ages preclude their hosting magma ocean planets, and we utilise them solely for the IMF comparison.

In our analysis of the magma ocean detection probabilities, we will focus on the three YMGs for which \citet{Bonati19} provide their complete simulation data; $\beta$~Pic, TW~Hyd and $\eta$~Cha. Furthermore, \citet{Bonati19} focus on detecting magma oceans around three different host star spectral types; A-, G- and M-stars. In Table~\ref{obs_data} we show the total number of stars in the \citet{Gagne18a,Gagne18b} data for these YMGs, as well as the total numbers of A-, G- and M-stars, and the latest observed spectral type in each YMG.  

\subsection{Monte Carlo simulations}

For our simulated YMGs we draw stellar masses from a \citet{Maschberger13} IMF, which has a probability distribution of the form
\begin{equation}
p(m) \propto \left(\frac{m}{\mu}\right)^{-\alpha}\left(1 + \left(\frac{m}{\mu}\right)^{1 - \alpha}\right)^{-\beta},
\end{equation}
where $\mu = 0.2$\,M$_\odot$ is the average stellar mass, $\alpha = 2.3$ is the \citet{Salpeter55} power-law exponent for higher mass stars, and $\beta = 1.4$ describes the slope  for low-mass objects. We sample this distribution in the mass range 0.1 -- 50\,M$_\odot$. 

For each YMG, we sample $N_{\rm stars}$ from this IMF, and repeat the process ten times to  gauge the stochasticity of randomly sampling this distribution. The choice of ten for the number of times we sample the IMF does not have any particular physical motivation, other than there are of order ten YMGs within 100\,pc, and one could imagine that -- if the IMF is universal -- sampling the IMF ten times would reproduce the observed numbers of stars in each YMG at least once.

  However, it is possible that sampling such low numbers of stars (e.g.\,\,44 in the case of $\beta$~Pic) could mean that our results are dominated or biased by a sampling error \citep[e.g.][]{Sarndal92}. To determine whether our results are affected by this, in the Appendix we present results where we sample the IMF 1000 times (instead of ten times) for each YMG and find that whilst the range of possible values increases, the overall results are very similar. We also repeat the experiment where we sample the IMF ten times, but change the random number seed ten times. Again, we find that our results are not dominated by a sampling error.

Each time, we count the numbers of A-stars \citep[defined as having masses in the range $1.5 - 3.0$\,M$_\odot$][]{DeRosa14}, G-stars \citep[defined as having masses in the range $0.8 - 1.2$\,M$_\odot$][]{Duquennoy91} and M-stars \citep[defined as having masses in the range $0.1 - 0.5\,$M$_\odot$][]{Fischer92}. Stars with masses outside these ranges (e.g.\,\,K- and F-) are not considered further in the analysis.

\subsection{Probability of detecting magma oceans}

We use the method and simulation data in \citet{Bonati19} to calculate the probability of detecting a magma ocean in each of $\beta$~Pic, TW~Hyd and $\eta$~Cha. Where our calculation (potentially) differs is in the numbers of A-, G- and M-stars used to determine the probability of detecting a magma ocean. For a full description of the method we refer the interested reader to \citet{Bonati19} but we provide a brief summary here.

The probability of detecting a magma ocean in a YMG, $P_{\rm MO}$ is given by
\begin{equation}
P_{\rm MO}(\lambda_{\rm cen}, d, \tau_\star, \epsilon) = 1 - \prod^{i=n_\star}_{i=1}\left(1 - \frac{\bar{n}_{\rm GI}\Delta t_{\rm MO}}{\Delta t_{\rm int}}\right), 
\label{pmo_detection}
\end{equation} 
where $n_\star$ is the number of stars of a given spectral type, $\bar{n}_{\rm GI}$ is the number of giant impacts that could be detected for a given wavelength $\lambda$, age of the YMG, $\tau_\star$, and the distance to the YMG, $d$. $\Delta t_{\rm MO}$ is the length of time a magma ocean planet would be detectable, and also depends on $\lambda$, $d$, $\tau_\star$ as well as the emissivity of the planet's atmosphere, $\epsilon$. $\Delta t_{\rm int}$ is the timescale for planet formation, which we keep fixed at 20\,Myr. 

The only variables in Eqn.~\ref{pmo_detection} are therefore the number of stars of a given spectral type, $n_\star$, the number of detectable giant impacts $\bar{n}_{\rm GI}$ (which is dependent on the age of the YMG; YMGs younger than 20\,Myr have more giant impacts) and the length of time over which a magma ocean planet would be observable, $\Delta t_{\rm MO}$. The latter two quantities are provided in the simulations of \citet{Bonati19}, and we summarise them in Table~\ref{sim_data}.

     

\begin{table}
  \caption{Data from \citet{Bonati19} used to calculate the probability of detecting a magma ocean around a star of a given spectral type at the distance of the YMG. We show the total number of giant impacts expected in a protoplanetary disc from $N$-body simulations, $n_{\rm GI, tot}$, the fraction of these giant impacts that would be detected with the 2.2$\mu$m filter on the ELT, $f_{\rm det}$, the total number of giant impacts that would therefore be detected, $\bar{n}_{\rm GI}$, and the length of time the magma ocean would be detectable with the ELT, $\Delta t_{\rm MO}$, assuming an atmospheric emissivity on the planet of 
$\epsilon = 0.01$.  }
  \begin{center}
    \begin{tabular}{cccccc}
      \hline
YMG & Sp. Type & $n_{\rm GI, tot}$ & $f_{\rm det}$ & $\bar{n}_{\rm GI}$ & $\Delta t_{\rm MO}$ \\
\hline
 & A & 5 & 0.75 & 3.75 & 1\,Myr \\ 
$\beta$~Pic & G & 5 & 0.78 & 3.9 & 0.7\,Myr \\ 
& M & 2 & 0.10 & 0.2 & 0.08\,Myr \\ 
       \hline
 & A & 16 & 0.70 & 11.2 & 0.4\,Myr \\ 
TW~Hyd & G & 20 & 0.60 & 12 & 0.08\,Myr \\ 
& M & 35 & 0.05 & 1.75 & 0.02\,Myr \\ 
       \hline
 & A & 16 & 0.38 & 6.08 & 0.05\,Myr\\ 
$\eta$~Cha & G & 20 & 0.32 & 6.4 & 0.01\,Myr\\ 
& M & 35 & 0 & 0 & 0.001\,Myr\\ 
       \hline
    \end{tabular}
  \end{center}
  \label{sim_data}
\end{table}

\section{Results}

\subsection{IMF sampling}

\begin{figure*}
 \begin{center}
\rotatebox{270}{\includegraphics[scale=0.7]{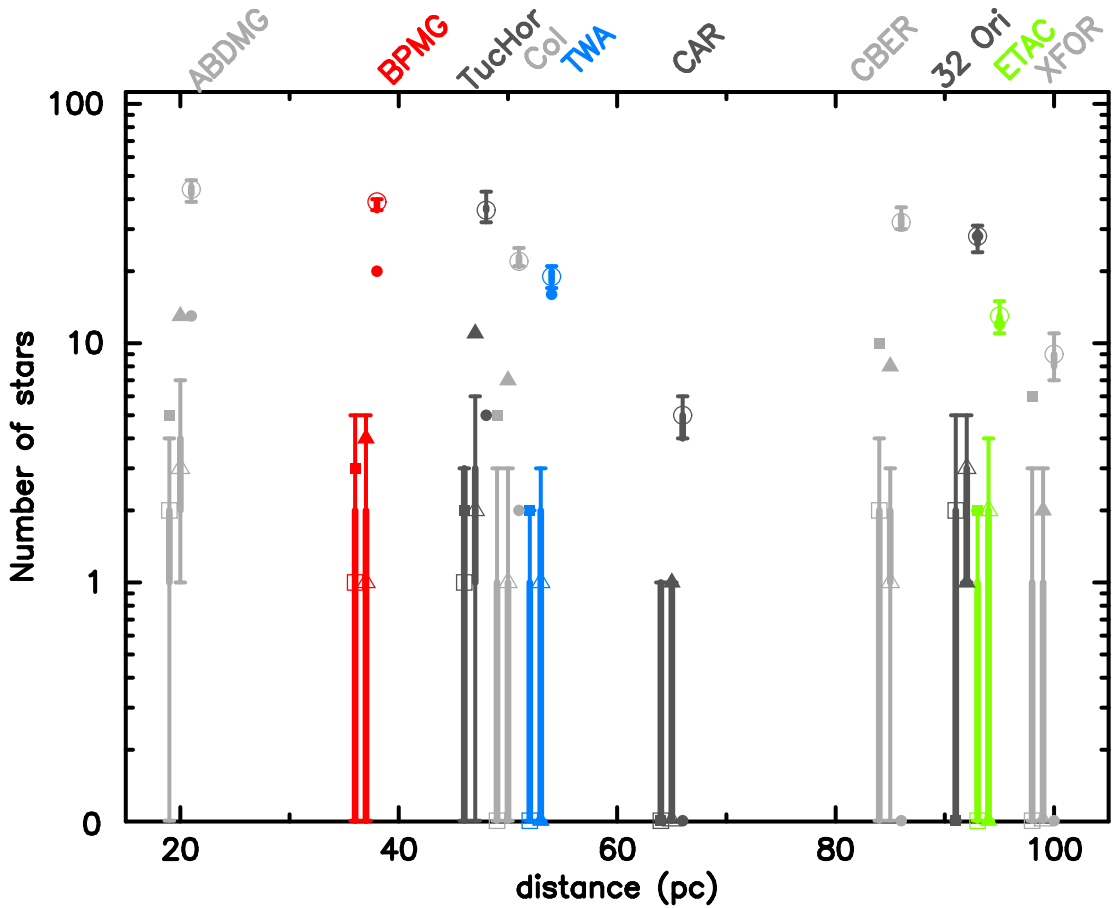}}
\caption{Number of stars by spectral type for nearby Young Moving Groups (YMGs). From left-to-right: AB~Dor moving group (ABDMG), $\beta$~Pic moving group (red symbols at a distance of 37\,pc, BPMG), Tucana Horologium moving group (TucHor), Columba moving group (Col), TW~Hydrae moving group (blue symbols at a distance of 53\,pc, TWA), Carina (CAR), Coma Bernices (CBER), 32~Orionis (32~Ori), the $\eta$~Cha moving group (green symbols at a distance of 94\,pc, ETAC) and the $\chi^1$~For moving group (XFOR). The filled symbols are the numbers of observed A-stars (squares), G-stars (triangles) and M-stars (circles) in each YMG. The open symbols are the median numbers of stars of the same spectral type from 10 random samplings of the IMF. The error bars indicate the full range of the number of stars of a given spectral type from sampling the IMF ten times and the thicker portions indicate the interquartile range. The different spectral types are offset for clarity. $\beta$~Pic, TW~Hyd and $\eta$~Cha are shown in different colours as we calculate the probabilities of detecting magma ocean planets in these three YMGs in Section~\ref{magma_detect}, with the other YMGs shown in two different shades of grey. }
\label{YMG_stochasticity}
\end{center}
\end{figure*}

We show the results of sampling the IMF ten times for each YMG in Fig.~\ref{YMG_stochasticity} (the results for sampling the IMF 1000 times are shown in Fig.~\ref{appendix:YMG_stochasticity}). The plot shows the numbers of A-stars (squares), G-stars (triangles) and M-stars (circles) for each YMG. The solid symbols are the observed numbers of stars in each spectral type for each YMG. The open symbols are the median number of stars in each spectral type from sampling ten realisations of the IMF and the error bars show the full range of values from these ten samplings. The thicker parts of the error bars indicate the interquartile range, which can be the same as the full range for  small numbers of total stars in a given YMG/spectral type). In Table~\ref{imf_data} we indicate for each YMG whether the numbers of A-, G- and M-stars are consistent with being drawn from a normal IMF.

For $\beta$~Pic, sampling an IMF up to a total number of 44 stars slightly underproduces the number of A- and G-stars compared to the observations (though still within the range from ten realisations of the IMF), but over-produces the number of M-stars by a factor of $\sim$2 (compare the filled circle, which is the total number of observed M-stars, to the open circle and its error bar, which is the maximum range of values from ten realisations of the IMF). Several authors \citep[e.g.][]{Gagne18b,Lee19a} have also noted this apparent deficit of M-stars in $\beta$~Pic compared to what would be expected from the IMF. 

Similarly, AB~Dor clearly has a deficit of M-stars, but has too many G- and A-stars based on randomly sampling the IMF. This is also the case for Coma Bernices, $\chi^1$~For, Tuc~Hor and Columba. 

In TW~Hydrae, sampling from a total of 22 stars reproduces the number of G-stars, but slightly underestimates the numbers of A- and M-stars. For $\eta$~Cha (the green symbols on the righthandside of the plot), the numbers of A-stars are underestimated from sampling the IMF, G-stars are overestimated but the number of  M-stars is roughly consistent with being drawn from the IMF.

Finally, for 32~Orionis  we would expect more A- and G-stars from sampling the IMF, but the number of M-stars is consistent with the IMF. In Carina (only 6 stars in total), the number of A- and G-stars are consistent with IMF sampling, but the absence of M-stars is inconsistent with the IMF, even with such low total numbers of stars.

In summary, of the 10 YMG in our chosen sample, all but two show a deficit of M-stars compared to a normal field-like IMF. Seven present an excess of A-stars (with only one showing a deficit), and six present an excess of G-stars (with two showing a deficit). As far as we are aware, ours is the first study to perform a comprehensive comparison with the IMF for all YMGs within 100\,pc.


\begin{table*}
  \caption{Comparison of the number of A-, G- and M-stars within each YMG in the dataset compiled by \citet{Gagne18a,Gagne18c} with the numbers obtained from randomly sampling the stellar IMF. If the YMG contains more stars of a spectral type than expected from sampling its total number of stars, $N_{\rm stars}$, we indicate this with a `$>$'. Conversely, if it contains fewer stars of a spectral type than expected from IMF sampling, we indicate this with a `$<$'. If the numbers are consistent with the value expected from the IMF, we indicate this with a `$=$'. The three YMGs for which we will calculate the probabilities for detecting magma ocean planets are listed first.}
  \begin{center}
    \begin{tabular}{cccccccc}
      \hline
YMG & $N_{\rm stars}$ & $N_{\rm A-stars, obs.}$ & IMF$_{\rm A-stars}$ & $N_{\rm G-stars, obs.}$ & IMF$_{\rm G-stars}$ & $N_{\rm M-stars, obs.}$ & IMF$_{\rm M-stars}$\\
     \hline	
     $\beta$~Pic & 44 & 3 & $>$ & 4 & $>$ & 20 & $<$\\
TW~Hyd & 22 & 2 & $>$ & 0 & $=$ & 16 & $<$ \\
$\eta$~Cha & 16 & 2 & $>$ & 0 & $<$ & 12 & $=$ \\
\hline
AB~Dor & 52 & 5 & $>$ & 13 & $>$ & 13 & $<$\\
Carina & 6 & 0 & $=$ & 1 & $=$ & 0 & $<$ \\
Tuc Hor & 43 & 2 & $=$ & 11 & $>$ & 5 & $<$ \\
Columba & 26 & 5 & $>$ & 7 & $>$ & 2 & $<$ \\
Coma~Ber & 38 & 10 & $>$ & 8 & $>$ & 0 & $<$ \\
32 Ori &  35 & 0 & $<$ & 1 & $<$ & 28 & $=$  \\
$\chi^1$~For & 11 & 6 & $>$ & 2 & $>$ & 0 & $<$ \\
       \hline
    \end{tabular}
  \end{center}
  \label{imf_data}
\end{table*}


There are three potential explanations for the deficit (or absence) of M-stars in the YMGs in our sample. First, it is possible  that the membership of these YMGs is incomplete, and we are missing the faintest members (i.e.\,\,M-stars). If the groups are incomplete, it is likely that analysis of the \emph{Gaia} Data Release 2 will help find further M-stars, or assign unconfirmed members to these YMGs. However, we note that the latest spectral type for each region (the final column in Table~\ref{obs_data}) often probes the M-/L-type regimes. For example, the latest type in $\beta$~Pic is an L7 object, so one may expect that stars brighter than this should have already been found and assigned to that group.

  A second explanation is that YMGs are the dynamically coherent remnants of star clusters that have undergone significant evolution and then disruption or dissolution. This could lead to a deficit of low-mass objects if the star clusters preferentially ejected M-stars during their early dynamical evolution. For this to be most efficient, the clusters would have to be extremely dense ($>10^4$\,M$_\odot$\,pc$^{-3}$) or be primordially mass segregated so that M-stars were located on the outskirts of the clusters and therefore less gravitationally bound to the cluster.

For populous star-forming regions containing more than 100 stars, it is possible to constrain the amount of dynamical evolution that has occurred by comparing the spatial structure to the relative local density of the most massive stars \citep[it is not possible to constrain the past density of a region using the present-day density, as a very dense region may have undergone rapid expansion, whereas a less dense region would have undergone less expansion,][]{Parker14b,Parker14e}. However, if the origin of the YMGs is from dense star clusters undergoing expansion, presumably we would also observe nearby dense star clusters `caught in the act' of dissolution in similar numbers to the numbers of YMGs, but such clusters do not appear to be common in the local solar neighborhood \citep{Bressert10}.

A third explanation is that YMGs are the outcome of star formation with a non-standard, top-heavy IMF. Whilst variations in the IMF have been suggested in extragalactic environments, and in the Galactic centre, there is little evidence for IMF variations close to the Sun \citep{Bastian10}, though see \citet{Dib18c}. If we sum the stars from all of these YMGs together, we have a total of 293 stars in the sample, 35 of which are A-type, 47 of which are G-type and only 96 are M-type. If we draw 293 stars from the IMF, we would expect at least 250 M-type stars. In fact, the problem is worse when summing together more than one distinct star-forming region, as the sum of many star-forming regions should result in a bottom-heavy `Integrated Galactic Initial Mass Function' \citep[IGIMF,][]{Kroupa03d}. Even if we ignore the IGIMF issue, in order to produce the summed total numbers of A- and G-stars, we would need to sample at least 1000 stars from a standard IMF, which would result in $\sim 850$ M-type stars. 
    

\begin{figure}
 \begin{center}
\rotatebox{270}{\includegraphics[scale=0.35]{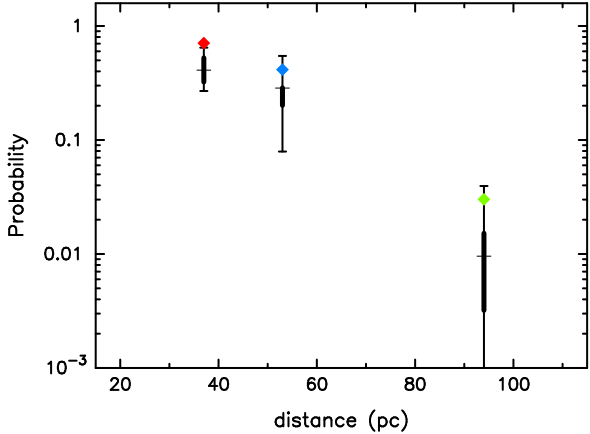}}
\caption{Probability of detecting a magma ocean planet with the ELT 2.2$\mu$m filter in the $\beta$~Pic (left), TW~Hydrae (centre) and the $\eta$~Cha (right) moving groups, assuming a planetary atmospheric emissivity of $\epsilon = 0.01$, using the method in \citet{Bonati19}. The diamond symbols indicate the probability of detecting a magma ocean planet around all M-stars, G-stars and A-stars combined, assuming the numbers of stars presented in recent literature   \citep{Gagne18a,Gagne18c}. The error bars indicate the full range of detection probabilities from randomly sampling the IMF ten times, and the thicker portions of the error bars indicate the interquartile range. The median values are shown by the horizontal lines.}
\label{magma_ocean_detect}
\end{center}
 \end{figure}

\subsection{Detecting magma ocean planets}
\label{magma_detect}

We now repeat several of the calculations performed by \citet{Bonati19} who quantified the probability for detecting magma ocean planets in nearby YMGs. We use the numbers of A-, G- and M-stars from \citet{Gagne18a,Gagne18c}, with estimates of the number of giant impacts that can be detected $\bar{n}_{\rm GI}$ in the ELT 2.22$\mu$m filter, and the timescale over which a magma ocean would be detected, $\Delta t_{\rm MO}$, again using the   ELT 2.22$\mu$m  filter and assuming an atmospheric emissivity of $\epsilon = 0.01$ (see Table~\ref{sim_data}). We use this emissivity and ELT filter as they represent the most optimal combination of planetary atmosphere and instrument for detecting magma ocean planets. However, we note that our results -- that stochastic sampling of the IMF can cause the detection probabilities to vary significantly -- would be relevant for other combinations of instrument/atmospheric conditions. We limit our analysis to $\beta$~Pic, TW~Hydrae and $\eta$~Cha because we only have access to the calculations for the length of time a magma ocean would be detectable for, $\Delta t_{\rm MO}$, for these three YMGs. However, we believe our results would be relevant to all of the YMGs listed in Table~\ref{obs_data}.

In Fig.~\ref{magma_ocean_detect} we show the probability of detecting magma ocean planets in $\beta$~Pic, TW~Hydrae and $\eta$~Cha, based on the currently confirmed members \citep{Gagne18a,Gagne18c} by the diamond symbols. Note that these values differ from those calculated in \citet{Bonati19} because we are using a more up-to-date census, with membership determined using {\it Gaia} DR 2 \citep{Gagne18a,Gagne18c}. Taking $\eta$~Cha as an example, \citet{Bonati19} use the data from \citet{Torres08} who report 1 A-star, 0 G-stars and 11 M-stars, whereas \citet{Gagne18a,Gagne18c} report 2 A-stars, 0 G-stars and 12 M-stars. \citet{Bonati19} calculate the probability of detecting a magma ocean planet in $\eta$~Cha with the ELT 2.2$\mu$m filter to be $\sim$0.01, whereas the additional A-star and M-star in our census increases the detection probability to 0.03 (the green diamond in Fig.~\ref{magma_ocean_detect}).

We then show the maximum range of this probability due to stochastic sampling of the IMF ten times\footnote{We show the results from sampling the IMF 1000 times, and the results for repeated sampling of the IMF ten times with different random number seeds in the Appendix.}, assuming the total number of confirmed members for each YMG.  The thicker portions of the error bars indicate the interquartile range, and the median values from the IMF sampling are indicated by the horizontal lines. Our motivation for this approach is to ask what the range of magma ocean detection probabilities could be due to sampling an IMF, and therefore what the range of detection probabilities would be for any hypothetical YMGs that may be newly discovered by e.g.\,\,Gaia~DR2, or later data releases \citep[e.g.][]{Liu20}, or if the membership of existing YMGs is significantly augmented by new detections \citep[e.g.][]{Binks20,Klutsch20}.

Taking TW~Hyd as an example (the central point), the probability for detecting a magma ocean planet based on its known membership is 0.41, but for the same number of stars, a YMG at this distance could have a magma ocean detection probability anywhere between 0.08 -- 0.55. 

Based on the observed numbers of A-, G- and M-stars in $\beta$~Pic, the probability of detecting a magma ocean planet is 0.71. However, because $\beta$~Pic appears to have a slight excess of both A- and G-stars, the detection probability based on the observed data lies above the maximum range predicted from randomly sampling the IMF (0.27 -- 0.65). In other words, we would expect the probability for detecting a magma ocean planet to be lower for a YMG at a similar distance and with a similar total number of stars, if those stars were drawn from from a more representative IMF than observed in $\beta$~Pic.

The number of A- and G-type stars in a star-forming region dominates the probability of detecting a magma ocean planet. A promising further option for detecting magma ocean exoplanets would be to observe the Sco Cen OB association with the ELT, which at a distance of 100 -- 150 pc \citep{deZeeuw99} hosts several hundred A-stars \citep{Mamajek02}. Using the same number of detectable giant impacts $\bar{n}_{\rm GI}$, and detectable magma ocean lifetime $\Delta t_{\rm MO}$ as $\eta$ Cham (which is at a similar distance to Sco~Cen), we would expect a detection probability of well over 90\% for 100 -- 200 A-stars.

\section{Conclusions}


We have performed Monte Carlo experiments to determine whether the numbers of stars (spectral types A, G and M) observed in YMGs are consistent with random sampling of the Galactic field IMF \citep{Maschberger13}, and what the range of expected values can be. We then determine the range of probabilities for detecting molten forming planets with future instrumentation in three YMGs ($\beta$~Pic, TW~Hydrae and $\eta$~Cha), and how this may be influenced by the stochastic nature of star formation. Our conclusions are the following:

(i) Eight of our sample of ten YMGs, including AB~Dor, $\beta$~Pic,  TW Hydrae, Tucana Horologium and Coma Bernices, appear to be deficient in M-stars compared to the Galactic field IMF \citep[something which has previously been noted in $\beta$~Pic,][]{Gagne18b}. Future data releases from \emph{Gaia} may add new members to these groups, although at present we cannot completely rule out an abnormal mode of star formation for these YMGs.

(ii) Seven of the YMGs also host more A-stars, and six host more G-stars  than would be expected from drawing the total number of stars in these YMGs from the field IMF. The probability of detecting magma ocean planets \citep{Bonati19} in $\beta$~Pic  is higher than would be expected if its IMF was field-like, due to the numbers of A- and G-stars being higher than expected from sampling the IMF.

(iii) Stochastic sampling of the IMF, which assumes that the star-forming molecular cloud fragments randomly to form stars, produces a significant spread in the expected numbers of planet-hosting stars, if the experiment is performed multiple (e.g.\,\,ten) times. This means that the probability of detecting magma ocean planets (and exoplanets in general) may vary between YMGs at similar distances and containing a similar number of stars.

\section*{Acknowledgements}

We thank the anonymous referee for their helpful report. We are also grateful to Tim Lichtenberg and Irene~Bonati for helpful discussions. RJP acknowledges support from the Royal Society in the form of a Dorothy Hodgkin Fellowship.


\newpage
\appendix

\section{IMF sampling}

When determining  the expected number of stars of a given spectral type, and the resulting probability of detecting magma oceans, we have stochastically sampled the IMF ten times (Figs.~\ref{YMG_stochasticity}~and~\ref{magma_ocean_detect}). However, the choice of ten is somewhat arbitrary, and in order to determine if our results could be affected by sampling bias \citep{Sarndal92}, we perform two further experiments.

  First, we repeat our calculations but sample the IMF 1000 times instead of ten. Secondly, we sample the IMF ten times, but repeat the same experiment with a different initial random number seed.

  \subsection{Sampling the IMF 1000 times}

\begin{figure}
 \begin{center}
\rotatebox{270}{\includegraphics[scale=0.7]{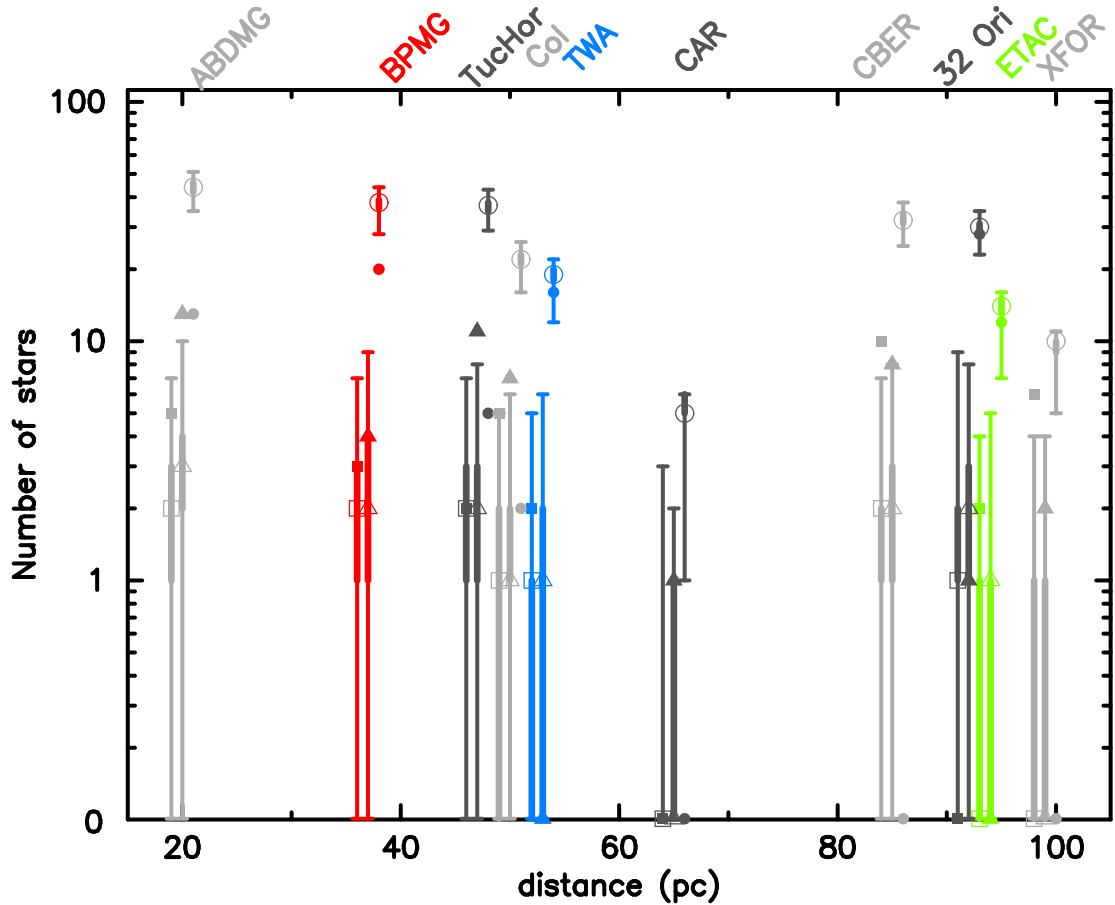}}
\caption{As Fig.~\ref{YMG_stochasticity}, but here we sample the IMF 1000 times instead of ten. Number of stars by spectral type for nearby Young Moving Groups (YMGs). From left-to-right: AB~Dor moving group (ABDMG), $\beta$~Pic moving group (red symbols at a distance of 37\,pc, BPMG), Tucana Horologium moving group (TucHor), Columba moving group (Col), TW~Hydrae moving group (blue symbols at a distance of 53\,pc, TWA), Carina (CAR), Coma Bernices (CBER), 32~Orionis (32~Ori), the $\eta$~Cha moving group (green symbols at a distance of 94\,pc, ETAC) and the $\chi^1$~For moving group (XFOR). The filled symbols are the numbers of observed A-stars (squares), G-stars (triangles) and M-stars (circles) in each YMG. The open symbols are the median numbers of stars of the same spectral type from 10 random samplings of the IMF.  The error bars indicate the full range of the number of stars of a given spectral type from sampling the IMF 1000 times. The different spectral types are offset for clarity. $\beta$~Pic, TW~Hyd and $\eta$~Cha are shown in different colours as we calculate the probabilities of detecting magma ocean planets in these three YMGs in Section~\ref{magma_detect}, with the other YMGs shown in two different shades of grey.}
\label{appendix:YMG_stochasticity}
\end{center}
\end{figure}

\begin{figure}
 \begin{center}
\rotatebox{270}{\includegraphics[scale=0.35]{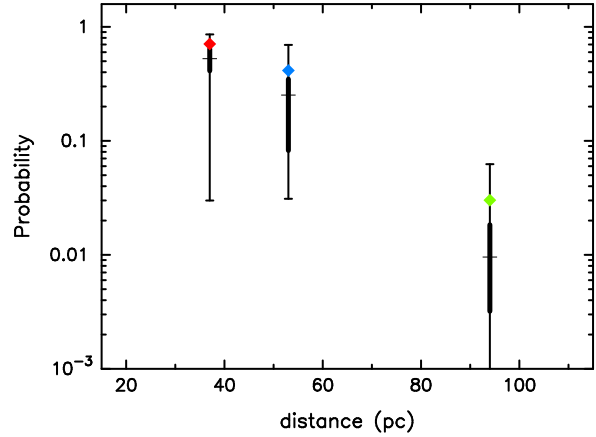}}
\caption{As Fig.~\ref{magma_ocean_detect},  but here we sample the IMF 1000 times instead of ten. We show the probability of detecting a magma ocean planet with the ELT 2.2$\mu$m filter in the $\beta$~Pic (left), TW~Hydrae (centre) and the $\eta$~Cha (right) moving groups, assuming a planetary atmospheric emissivity of $\epsilon = 0.01$, using the method in \citet{Bonati19}. The coloured diamond symbols indicate the probability of detecting a magma ocean planet around all M-stars, G-stars and A-stars combined, assuming the observed numbers of stars presented in recent literature \citep{Gagne18a,Gagne18c}. The error bars indicate the full range of detection probabilities from randomly sampling the IMF 1000 times, and the thicker portions of the error bars indicate the interquartile range. The median values are shown by the horizontal lines.}
\label{appendix:magma_ocean_detect}
\end{center}
\end{figure}

  When we sample the IMF 1000 times, our main results are essentially unchanged (Figs.~\ref{appendix:YMG_stochasticity}~and~\ref{appendix:magma_ocean_detect}). The range of possible values is slightly larger, as one would expect from sampling the IMF 1000 times instead of 10. However, despite this larger range, there is no change to the results for six of the ten YMGs ($\beta$~Pic, $\eta$~Cha, Carina, Tuc~Hor, 32~Ori and $\chi^1$~For). In AB~Dor the observed number of A-stars now  lies within the range of values from sampling the IMF, but the numbers of G-stars and M-stars are still inconsistent with random sampling. In Columba, the number of A-stars is now at the very top of the range expected from IMF sampling (i.e.\,\,this could be a 1/1000 event), though the numbers of G-stars and M-stars are still inconsistent with the IMF. Similarly, the number of G-stars in Coma Bernices is now reproduced by IMF sampling (again, at a 1/1000 level).

The only notable difference between sampling the IMF 10 or 1000 times is that we can reproduce the observed number of M-dwarfs in TW Hyd when sampling the IMF 1000 times, whereas we do not reproduce the observed number of M-dwarfs when sampling the IMF 10 times (compare the solid blue circular points in Fig.~\ref{YMG_stochasticity} with Fig.~\ref{appendix:YMG_stochasticity}). 

  We recalculate the probability of detecting magma ocean planets and show the results for 1000 samplings of the IMF in Fig.~\ref{appendix:magma_ocean_detect}. As expected from a larger number of samplings, the full range of possible values (depicted by the full `whisker' error bars) has increased, as rarer outcomes occur. The interquartile ranges are also larger, but the median values are similar to those for ten samplings of the IMF. Most notably, because of the wider range of possible values, the probability of detecting a magma ocean around a star in $\beta$~Pic using the observed census (the red diamond symbol) now lies within the parameter space of the simulations, rather than just above the highest value when we sample the IMF ten times (compare the position of the red diamond symbol in Fig.~\ref{appendix:magma_ocean_detect}~versus~Fig.~\ref{magma_ocean_detect}). 

  \subsection{Repetitive sampling}

\begin{figure*}
  \begin{center}


   \rotatebox{270}{\includegraphics[scale=0.25]{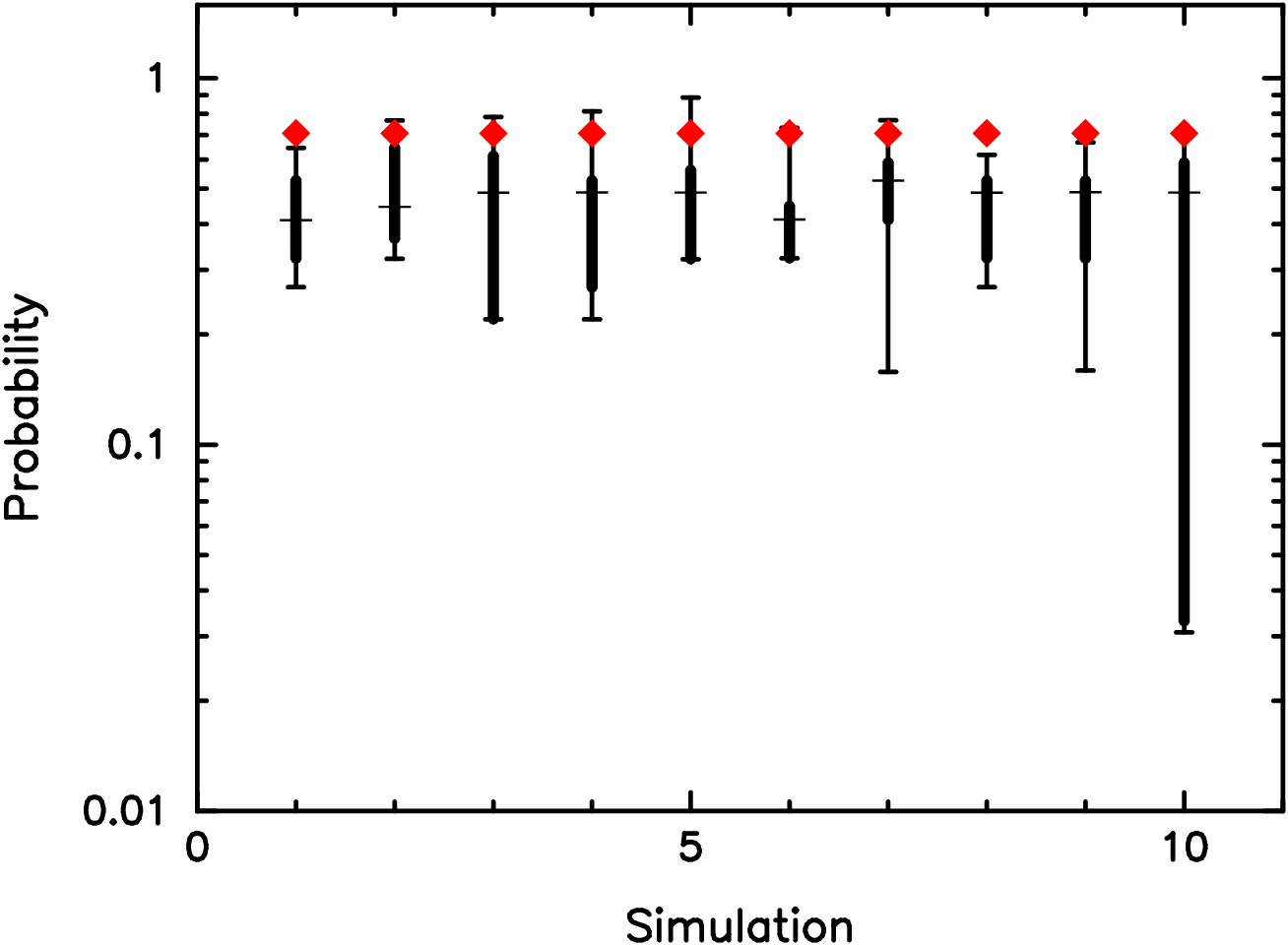}}
   \hspace*{0.2cm}
   \rotatebox{270}{\includegraphics[scale=0.25]{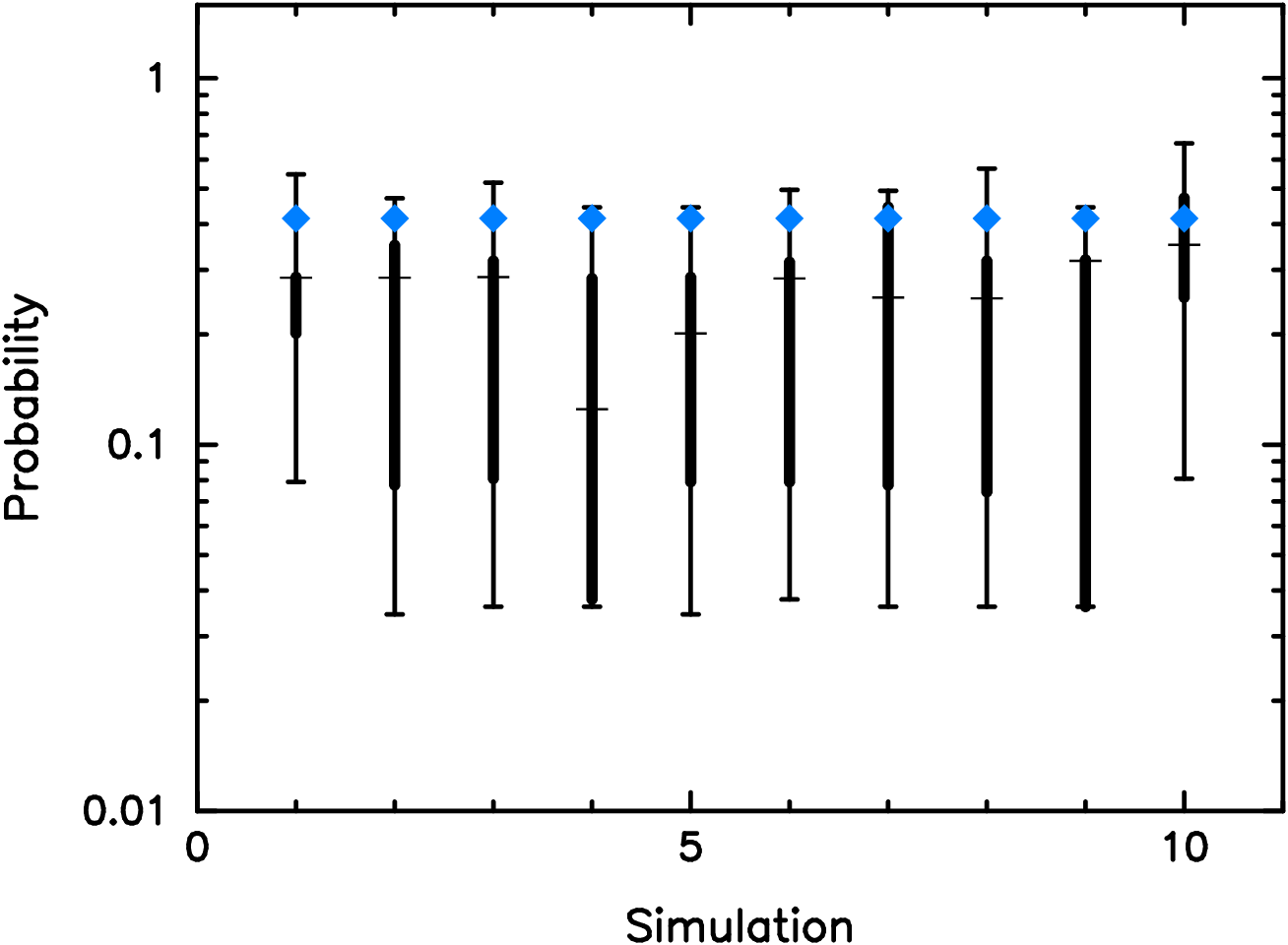}}
   \hspace*{0.2cm}
   \rotatebox{270}{\includegraphics[scale=0.25]{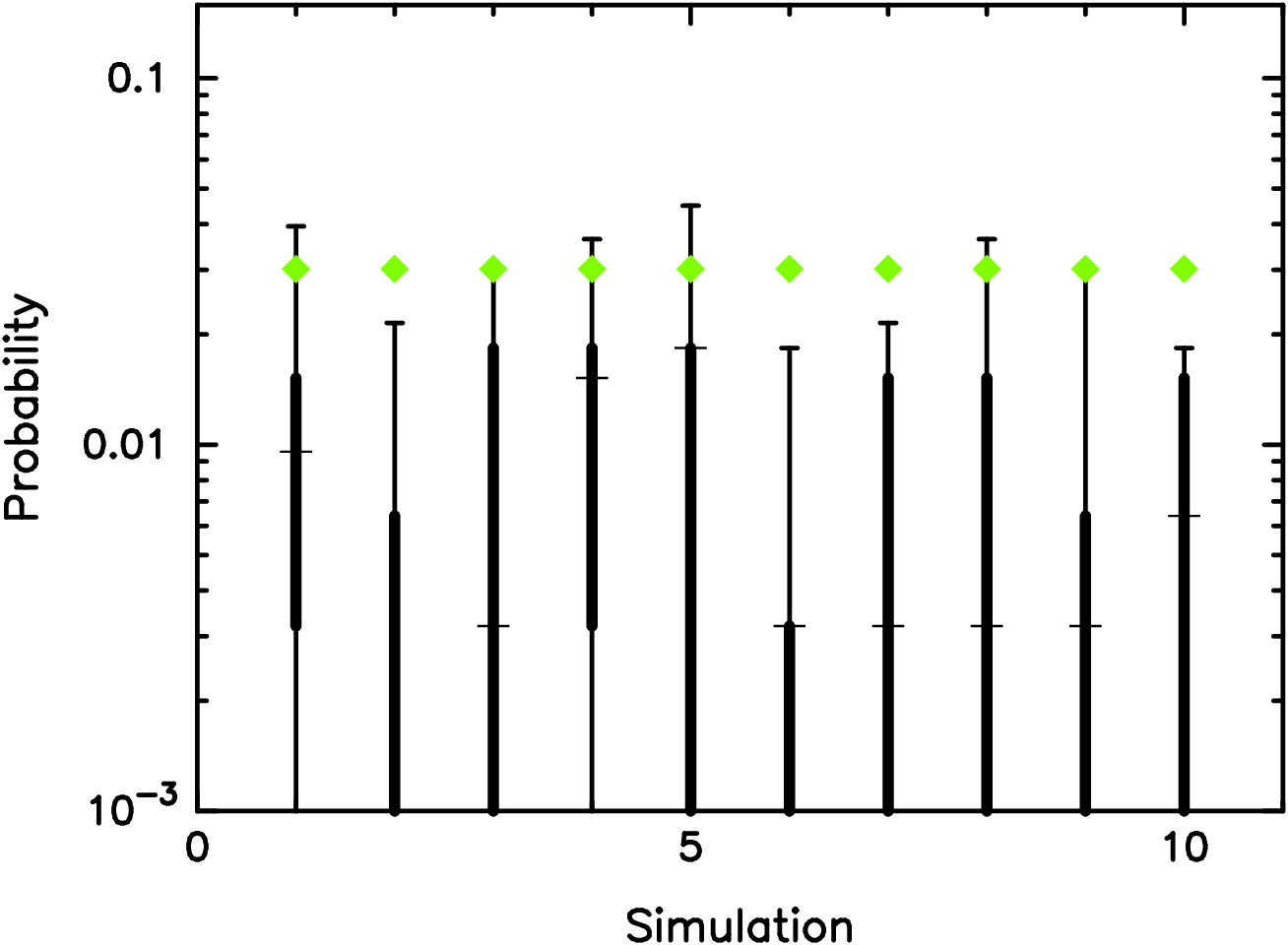}}
\caption{Probability of detecting a magma ocean planet with the ELT 2.2$\mu$m filter in the $\beta$~Pic (left), TW~Hydrae (centre) and the $\eta$~Cha (right) moving groups, assuming a planetary atmospheric emissivity of $\epsilon = 0.01$, using the method in \citet{Bonati19}. The coloured diamond symbols indicate the probability of detecting a magma ocean planet around all M-stars, G-stars and A-stars combined, assuming the numbers of stars presented in recent literature  \citep{Gagne18a,Gagne18c}. The error bars indicate the full range of detection probabilities from randomly sampling the IMF 10 times, and the thicker portions of the error bars indicate the interquartile range. The median values are shown by the horizontal lines. In each figure panel, the experiment has been repeated ten times and the original datapoint from Fig.~\ref{magma_ocean_detect} is on the far left of each panel. Note the different y-axis scale in panel (c).}
\label{appendix:box_whisker_all}
\end{center}
 \end{figure*}

  We now revert back to sampling YMGs from the IMF ten times, but change the random number seed used to initialise our Monte Carlo simulations and conduct the experiment ten times. To avoid producing an unreadable plot, we do not show the results for the numbers of A-, G- and M-stars for all ten YMGs in (as shown in Fig.~\ref{YMG_stochasticity}). However, as these numbers directly contribute to the probability of detecting magma ocean planets, any potential sampling error should be present in these magma ocean detection probabilities.  We show the box and whisker plots from Fig.~\ref{magma_ocean_detect} in separate figure panels in Fig.~\ref{appendix:box_whisker_all} (note the change in axes for the righthand panel, which shows the results for $\eta$~Cha).

  In all panels of Fig.~\ref{appendix:box_whisker_all}, the leftmost box and whisker plot is the same as that in Fig.~\ref{magma_ocean_detect}. There is very little difference between the results when repeating the experiment ten times. The probability of detecting a magma ocean in $\beta$~Pic using the observed population of stars lies within the simulation range around 50\,per cent of the time, and interestingly, the corresponding value for  $\eta$~Cha falls outside of the simulation range 50\,per cent of the time (having been within the simulation range in Fig.~\ref{magma_ocean_detect}). The median probability of detecting a magma ocean planet is almost constant for $\beta$~Pic and TW~Hyd, with a lot more scatter for the more distant $\eta$~Cha. \\

\subsection{Summary}
  
  Overall, our results suggest that sampling errors do not affect our conclusions. The probability of detecting a magma ocean planet is still highly dependent on stochastic sampling of the stellar IMF, and the probability can vary by a factor of at least two.


\bibliography{general_ref}
\bibliographystyle{aasjournal}



\end{document}